# Desiderata for a biomedical knowledge network: opportunities, challenges and future Directions

Chunlei Wu[1,*], Hongfang Liu[2], Jason Flannick[3], Mark A. Musen[4], Andrew I. Su[1], Lawrence Hunter[5], Thomas M. Powers[6], Cathy H. Wu[7,*]

[1]Department of Integrative Structural and Computational Biology, The Scripps Research Institute, La Jolla, CA 92037, United States.

[2]The McWilliams School of Biomedical Informatics, University of Texas Health Science Center at Houston, Houston, Texas 77030, United States.

[3]Programs in Metabolism and Medical and Population Genetics, the Broad Institute of MIT and Harvard, Cambridge, MA 02142, United States.

[4]Center for Biomedical Informatics Research, Stanford University, Palo Alto, CA 94304, United States.

[5]Department of Pediatrics, University of Chicago, Chicago, IL 60637, United States.

[6]Department of Philosophy, University of Delaware, Newark, DE 19716, United States

[7]Department of Computer and Information Sciences, University of Delaware, Newark, DE 19716, United States.

*Corresponding authors. Department of Integrative Structural and Computational Biology, The Scripps Research Institute, La Jolla, CA 92037, United States. E-mail: cwu@scripps.edu. Department of Computer and Information Sciences, University of Delaware, Newark, DE 19716, United States. E-mail: wuc@udel.edu.

**Abstract**
Knowledge graphs, collectively as a knowledge network, have become critical tools for knowledge discovery in computable and explainable knowledge systems. Due to the semantic and structural complexities of biomedical data, these knowledge graphs need to enable dynamic reasoning over large evolving graphs and support fit-for-purpose abstraction, while establishing standards, preserving provenance and enforcing policy constraints for actionable discovery. A recent meeting of leading scientists discussed the opportunities, challenges and future directions of a biomedical knowledge network. Here we present six desiderata inspired by the meeting: (1) inference and reasoning in biomedical knowledge graphs need domain-centric approaches; (2) harmonized and accessible standards are required for knowledge graph representation and metadata; (3) robust validation of biomedical knowledge graphs needs multi-layered, context-aware approaches that are both rigorous and scalable; (4) the evolving and synergistic relationship between knowledge graphs and large language models is essential in empowering AI-driven biomedical discovery; (5) integrated development environments, public repositories, and governance frameworks are essential for secure and reproducible knowledge graph sharing; and (6) robust validation, provenance, and ethical governance are critical for trustworthy biomedical knowledge graphs. Addressing these key issues will be essential to realize the promises of a biomedical knowledge network in advancing biomedicine.

# 1 Introduction

Knowledge graphs (**KG**s) have become critical tools for semantic integration of biomedical data in an increasingly computable, inference-bearing and explainable knowledge system, allowing researchers to uncover meaningful patterns, predict new relationships, and extract insights to enable biomedical knowledge discovery. To explore the opportunities, challenges and future directions of a biomedical knowledge network (as a collection of individually built KGs), we hosted a "Knowledge Networks in Biomedical and Behavioral Research" meeting on May 28-29, 2025 (KN meeting 2025), organized under the auspices of the Knowledge Graph Working Group of the US National Institutes of Health (NIH). The meeting brought together leading scientists in the field from multiple NIH-funded programs such as the Biomedical Data Translator (Fecho et al. 2025) and Common Fund Data Ecosystem (CFDE n.d.) as well as the NSF partnership program, Proto-OKN (Proto-OKN n.d.), with a total of 88 participants including NIH program officials.

The goals of the meeting were multifold: (1) to identify opportunities and challenges in KG research; (2) to discuss strategies to facilitate developments in AI technologies and approaches for building KGs; and (3) to develop guidelines for ethical, transparent, and informed reuse of data, knowledge, and models (KN meeting 2025). Accordingly, we organized three half-day panel sessions to address each of the topics with breakout sessions for discussion. This perspective article presents our collective observations and thoughts, organized as six (6) desiderata, inspired by the keynote and panel presentations and vibrant discussions from all meeting participants. We summarize several challenges as well as opportunities and future directions from the desiderata for a biomedical knowledge network.

# 2 Desiderata

Realizing the full potential of a biomedical knowledge network requires a principled fusion of domain specificity, semantic rigor, computational scalability, reproducibility and trustworthiness. By addressing the current challenges and harnessing strategic opportunities as detailed in the desiderata, we aspire to build a biomedical knowledge network capable of supporting robust, interpretable, and actionable biomedical discovery (**Figure 1**).

**Desideratum 1. Inference and reasoning in biomedical knowledge graphs need domain-centric approaches**

Effective inference within biomedical knowledge graphs demands methodologies that are attuned to the semantic and structural complexities of biomedical data. Unlike general-purpose KGs, biomedical KGs exhibit high relational density, many-to-many topologies, and a mix of symmetric and directed relationships (Stear et al. 2024). Edge types often range from simple assertions to qualified statements, carrying metadata such as confidence scores, provenance, or temporality. Many applications require reasoning over complex subgraphs, or finding semantically constrained long paths. These characteristics render conventional edge prediction techniques inadequate. In this domain, inference is not merely a mathematical task but an inherently semantic one, calling for novel approaches that leverage graph structure through logical and heuristic inference, tailored to domain-centric application needs.

To harmonize symbolic reasoning (e.g., logic- and ontology-based methods) with domain-centric approaches, biomedical inference engines must operate across heterogeneous data scales, from biological centric genes and variants to clinical centric patient data, while preserving provenance, enforcing policy constraints, and accommodating both dynamic updates and partial knowledge. Critically, methods must respect the semantics of node and edge types (e.g., *part-of*, *regulates*, *expressed-during*) to ensure biologically plausible information flow and interpretable reasoning pathways. The Data Distillery Knowledge Graph (DDKG), built within the NIH Common Fund ecosystem, exemplifies how semantic reasoning can be operationalized at scale (DDKG n.d.). Extending the Unified Biomedical Knowledge Graph (UBKG n.d.), DDKG employs ontologies, graph motifs, and structured reasoning pathways to integrate datasets such as GTEx, LINCS, HuBMAP, and Kids First. It enables use cases in disease mechanism exploration, therapeutic target identification, and developmental biology, highlighting the potential of reasoning-ready KGs to drive translational insights.

Looking forward, next-generation inference engines must address several open challenges: **a) multimodal integration** should be enabled to incorporate diversified causal priors and spatially explicit data (e.g., anatomical atlases such as HuBMAP and Allen Brain); **b) temporal reasoning** is a necessity when inferencing across developmental stages or disease trajectories; **c) reasoning over large subgraphs and finding semantically constrained long paths** (e.g. representations of complex regulatory networks), not just edge prediction or retrieving triples; **d) cross-species alignment**, building on efforts like UPheno (Matentzoglu et al. 2024) to support translational and comparative studies. **e) uncertainty representation** is especially important in domains like dose-dependent gene expression (e.g., LINCS), where biological variability must be modeled with nuance; **f)** finally, **the balance between abstraction and fidelity** must be managed. Driven by the specific use cases, practitioners need tools that support fit-for-purpose KG abstraction while retaining links to detailed source data. Such tools should enable dynamic reasoning over large, evolving graphs without requiring costly retraining.

In summary, realizing the full potential of inference in biomedical KGs requires a principled fusion of semantic rigor, computational scalability, and domain specificity. By grounding machine reasoning in biological meaning, we can build systems capable of supporting robust, interpretable, and actionable biomedical discovery.

**Desideratum 2. Harmonized and accessible standards are required for knowledge graph representation and metadata**

The core methods for representing and building knowledge graphs have changed little since AI researchers first developed semantic networks in the 1960s. These early networks enabled applications like intelligent tutoring systems and offered flexibility and intuitive structure, but they lacked standardized representations. The absence of shared formalisms led to fragmented approaches that hindered interoperability. While some efforts, such as DARPA's 1990s Knowledge Sharing Effort (Patil et al. 1992), attempted to bridge these gaps, their impact remained modest and limited within research laboratories. A turning point came in the 2000s, with the introduction of Web standards—RDF (Resource Description Framework) and OWL (Web Ontology Framework)—endorsed by the World Wide Web Consortium. These standards enabled scalable creation of linked data graphs and reusable ontologies, driving a renaissance in KG development, coupled with growing attention to data governance. Today, the landscape is again fragmented, with the recent surge of diverse graph database systems, such as labelled property graphs, now competing commercially with RDF. The evolution of the underlying graph database technology was typically driven by the increasing demands of scalable querying and dissemination, nevertheless, the need for standards and their harmonization remains central to advancing KGs at scale.

Several key requirements are broadly applicable to the current process of building a biomedical knowledge networks.

**Standards are essential.** The resurgence of the KG community has been driven by the emergence of standards. While biomedical domain specific standards exist for graphs and ontologies, they remain incomplete. More accessible and expressive formalisms are needed to model *complex relationships*, *contextual information*, and *provenance*—capabilities that labeled property graphs support, albeit in an *ad hoc* manner. Critically, there is still a lack of widely accepted standards for *graph metadata*—descriptors that would facilitate the discovery, reuse and integration of graph content. There is an urgent need for structured reporting guidelines for annotating graphs and graph fragments, analogous to existing standards for genomic datasets such as MIAME (Brazma et al. 2001).

There is a **growing demand for standardized ontologies, value sets, and other curated semantic resources.** KGs are most effective when they use interoperable, well-defined terms linked to established ontologies. To enable scalability, there must be sustained investment in both the maintenance of existing ontologies and the development of new ones for scientific domains that remain underrepresented. Although many biomedical ontologies are actively curated and well supported, continued investment is essential to build and sustain a comprehensive biomedical knowledge network. In parallel, there must be a commitment to preserving these semantic resources in open, publicly accessible repositories such as BioPortal (Whetzel et al. 2011).

Of critical importance are **standards across all layers of the KG building ecosystem**: the ontologies that define terms, the graph representations themselves, the metadata that describes them, and the tools used to build, visualize, and integrate them. Much of what we have learned from standardizing ontologies and software systems can now inform the development of standards for KGs. However, the scale of the challenge is far greater. There are orders of magnitude more content, and the necessary graph models, ontologies, and supporting tools to realize this vision have yet to be developed. The Biolink model (Unni et al. 2022), as used in the Translator project, provides a compelling example of the direction and potential needed to meet this challenge.

In summary, sustained progress in biomedical KGs depends on harmonized, accessible standards for graph representations, ontologies and metadata. Addressing current gaps will require coordinated investments in standard development, resource maintenance, and tool interoperability across the entire ecosystem.

**Desideratum 3. Robust validation of biomedical knowledge graphs needs multi-layered, context-aware approaches that are both rigorous and scalable.**

Validation, in this context, is not merely a technical checkpoint but a continuous, context-sensitive process that must evolve in parallel with the knowledge it represents. Constructing domain-specific knowledge graphs often relies on expert curation, integrating evidence from literature, real-world data, and other sources. Yet biomedical knowledge is constantly shifting, with new findings frequently revising or contradicting earlier ones. Consequently, validation strategies must go beyond verifying the existence or direction of relationships to assess their strength of evidence, context specificity, and potential modifiers.

As biomedical KGs are continuously updated to reflect new discoveries, evolving terminologies, and shifting clinical guidelines, they face challenges in data heterogeneity, scalability, standardization, and entity resolution. Quality issues, such as incorrect hierarchical relationships or missing terminology mappings, can lead to flawed clinical insights, suboptimal cohort selection, and diminished analytic reproducibility.

Effective quality management involves both assessment (e.g., evaluating completeness, consistency, timeliness, and accuracy) and improvement (e.g., through error detection, correction, and gap completion). Existing methods span manual curation, rule-based systems, statistical checks, machine learning, and large language model (LLM)-augmented approaches. However, the absence of a consensus definition of "fitness for use" hampers automation, benchmarking and reuse. Of equal importance is the ability to track "knowledge drift", changes in structure, semantics, or provenance that affect downstream interpretations. Systematic provenance tracking should accompany each assertion, detailing its source, method of extraction, and evidence type.

Validation must also **represent uncertainty at multiple levels**, across nodes, edges, and subgraphs. This includes marking conflicting evidence, tagging causal claims with strength and direction, and maintaining version control in dynamic graphs.

Critically, validation should be **use-case driven**. For instance, tolerance for ambiguity or error differs substantially between applications in information retrieval, diagnostic inference, and causal modeling. Validating graph construction is distinct from validating graph utility; each requires tailored metrics, interpretive frameworks, and benchmarks. Rather than relying on a single gold standard, task-specific validation protocols must be developed. Initiatives such as TRIPOD-AI (Collins et al. 2021), which provide reporting standards for predictive models, could inform similar frameworks for KGs, defining minimum requirements for transparency, evidence sufficiency, and explainability. Benchmark datasets, checklists, and reference implementations are needed to promote reproducibility and cross-institutional adoption.

Additionally, **ontological mappings present another layer of complexity**. Inaccurate or ambiguous mappings can compromise interoperability, especially when mappings conflate related concepts (e.g., "broader-than" vs. "equivalent-to"). Mapping provenance and confidence scores must be made explicit, with fine-grained metadata to support auditability, trust and downstream utility. Standardizing such practices across KGs is a critical next step.

Moving forward, **community-wide collaboration is essential**. Key priorities include: a) developing shared QA pipelines that incorporate provenance, uncertainty tagging and confidence scoring; b) building tools for dynamic validation that accommodate evolving knowledge and contextual shifts; c) supporting retrospective validation (e.g., validating past predictions with future outcomes; and d) establishing standards for edge-level metadata, ontological mappings, and equivalence annotations.

The central challenge is defining and operationalizing validation frameworks that are scalable, interpretable, and aligned with scientific rigor. While peer review of KGs is valuable, it is currently too labor-intensive to scale. Semi-automated pipelines and benchmarked validation tools are essential for sustainable KG development.

In summary, validation is not an endpoint but an iterative, multidimensional process embedded throughout the lifecycle of biomedical KGs. Embedding provenance, context, and shared standards at every level is essential to ensure these systems are not only computationally robust but also trustworthy and fit-for-purpose.

**Desideratum 4. The evolving and synergistic relationship between knowledge graphs and large language models is essential in empowering AI-driven biomedical discovery.**

In the era of powerful large language models (LLMs), knowledge graphs remain essential. While LLMs excel at interpreting unstructured data and generating natural language outputs, KGs offer structured, precise, and provenance-rich representations that help constrain LLM hallucinations, support explainability, and ensure scientific rigor. Encoding biomedical knowledge as graphs inherently imposes a degree of formalization that benefits both computational inference and human inspection. In this sense, KGs act as a source of truth for LLMs, grounding inference and generated content within biomedically sensible context.

KGs and LLMs thus play complementary roles: KGs provide rigorously defined representations, identifiable provenance, and the potential for logical and semantic reasoning, while LLMs are adept at processing, generating, and transforming unstructured input. LLMs can assist KG construction by extracting triples from text or identifying inconsistencies (Pop et al. 2025), while KGs can enhance LLMs outputs through retrieval-augmented generation (RAG) or by injecting validated facts into prompts. Compared to LLMs, KG-based reasoning is more "human-comprehensible" and transparent, allowing domain experts to inspect and refine knowledge directly, which can be regarded as "living textbooks of the future" for biomedicine.

Several promising integration patterns are emerging: **a) graph-augmented retrieval** to ground LLM responses in a biomedical domain-specific context; **b) LLM-assisted curation** to support semi-automated graph construction; **c) hybrid reasoning**, blending logical inferences from KGs with LLM-generated hypotheses or narratives. LLMs can also improve KG usability by acting as natural language interfaces, allowing experts to query graphs conversationally and receive contextual, structured responses—lowering the cognitive barrier to interacting with complex KGs encoded in triple format.

Despite promise, challenges remain. Current approaches often fall short in terms of scalability, utility, and reliability. Human-in-the-loop workflows are critical for validating LLM-generated triples, and ensuring contextual accuracy, especially in high-stakes domains like precision medicine. Novel methods are needed to scale expert oversight while preserving scientific fidelity. A key opportunity is to use LLMs not as autonomous knowledge creators, but as assistive agents in structured curation workflows, akin to code review assistants in software development. For instance, projects like Uberon have successfully adopted semi-automated, expert-guided ontology development (Haendel et al. 2014). This model combines LLM efficiency with human judgement, aligning automation with domain-specific epistemic constraints.

Mitigating hallucination and ensuring provenance are also critical in KG-LLM integration. Strategies include prompt engineering, enforcing structured outputs (e.g., triple formats), citation tagging, and external validation against trusted datasets or knowledge bases. Clinical use cases also require an extra layer of privacy safeguards and access controls.

In summary, KGs provide the structure, provenance, and semantic rigor that LLMs lack, while LLMs offer new ways to build, maintain, and interact with KGs. Together, they can amplify human expertise and enable biomedical AI systems that are powerful, transparent, and trustworthy—provided the field invests in validation, standards, and human-centered tooling to realize their full potential.

**Desideratum 5. Integrated development environments, public repositories, and governance frameworks are essential for secure and reproducible knowledge graph sharing**

Knowledge graphs are fundamentally about knowledge sharing and reuse, the power of the graph-based inference and reasoning arises from the broad scope of integrated knowledge. While the projects relying on biomedical KGs have rapidly increased, the development and sharing of biomedical KGs are hindered by the **lack of robust development environments** for standards-based graph construction. An important aspect of a reproducible KG is the ability to share the build process, not just the resulting graph (Callahan et al. 2024). New development environments that facilitate graph construction best practices and reproducibility will be broadly impactful. While commercial graph databases offer user-friendly interfaces and end-to-end tooling, RDF-based graph engineering remains a fragmented and technically demanding process. Efficient tools for building, visualizing, debugging, and maintaining large, standards-compliant graphs are urgently needed. Such environments would not only accelerate development but also enhance reproducibility, rigor, and interoperability across graph-based applications.

Equally critical is the creation of **public, searchable repositories for KGs**. Currently, there is no recognized archive where researchers can publish, explore, and reuse KG content. A well-designed repository would support graph visualization, metadata-rich indexing, and modular reuse—allowing developers to merge existing graphs by value or by reference into their own projects. These repositories should align with existing biomedical infrastructure (e.g., BioPortal, UMLS) and adopt metadata standards for graph description and provenance. Just as the NIH endorses standardized value sets, it could similarly support a set of reference graphs to promote reuse and quality assurance.

Building on NIH's 2023 Data Management and Sharing Policy ("Data Management and Sharing Policy" n.d.), a parallel **Knowledge Management and Sharing Policy** could be developed. This would include expectations for source transparency, build reproducibility, licensing disclosure, update frequency, and provenance documentation. Automated rebuild pipelines, standardized graph packaging, and repository-based dissemination could significantly improve trust and sustainability. Addressing licensing challenges, often a roadblock for reuse, should be a priority in any future governance framework.

In summary, to ensure that biomedical KGs are secure, reproducible, and shareable at scale, the field must invest in purpose-built development environments, curated public repositories, and governance models that balance openness with ethical responsibility. These infrastructure and policy investments will be essential to support community-driven, trustworthy knowledge sharing in the era of AI-enhanced biomedical discovery.

**Desideratum 6. Rigorous validation, provenance, and ethical governance are critical for trustworthy biomedical knowledge graphs**

Biomedical knowledge graphs are complex, evolving systems that integrate diverse and possibly sensitive information. To be trustworthy and impactful, they must be validated rigorously, governed transparently, and built on provenance-rich foundations. Yet, the scale, complexity, and domain-specific nature of biomedical knowledge pose significant challenges. Biomedical KGs differ from other domains due to their rapid knowledge evolution, uncertainty in evidence, and the sheer diversity of entity and relationship types, such as genes, proteins, phenotypes and clinical concepts.

Validation remains a persistent challenge. Unlike static datasets, KGs require continuous quality assessment as knowledge evolves. This includes verifying not just the presence of connections, but the context, evidence strength, and potential modifiers. Manual validation is costly and resource-intensive; scalable solutions must combine automation with human oversight. Existing automated approaches to support the reuse of biomedical data, such as the (Re)usable Data Project (Carbon et al. 2019), provide useful frameworks, but broader adoption will require new methods, including checklists for licensing and provenance, and smarter automation that can support expert review.

The advantage of KGs over black-box AI models like LLMs is their inherent transparency: **the provenance of every statement can, in principle, be traced**. However, this ideal is not always realized. Current frameworks (Stear et al. 2024; Putman et al. 2024; Callahan et al. 2024) emphasize automated graph construction and build pipelines, including source and mapping metadata for each node and edge. While this represents progress, challenges persist—especially in handling ambiguous ontology mappings, context-sensitive validity, and the articulation of design intent, audience, and use case.

**Technical barriers are compounded by governance gaps**. Licensing restrictions often inhibit reuse, derivation, or redistribution of data, with many data sources having unclear terms (Carbon et al. 2019). Ethical concerns also demand renewed attention. While some issues—like re-identification risks—are familiar, the structured nature of KGs introduces new potential harms, such as emergent inferences or exposure of sensitive linkages. Governance must also address equity and sovereignty, particularly for marginalized groups such as Indigenous communities, where control over data use is paramount.

**Privacy must be balanced with openness**. The principle of "as open as possible, but as restricted as necessary" (Landi et al. 2020) adheres with more general principles of conducting open and careful science (Resnik 1998). Under this principle, technical strategies—homomorphic encryption, differential privacy, and privacy-preserving embeddings—offer potential for fine-grained control without undermining usability. Capturing and honoring data-use preferences from participants is both an ethical imperative and a technical challenge. The security of KGs must be protected against misuse and adversarial exploitation. It will be important that KG security practices conform with international standards, e.g. the ISO 27001 requirements on information security management (ISO 27001 2022) and relevant law such as the EU's General Data Protection Regulation (GDPR 2018).

In summary, to build a trustworthy biomedical knowledge network, the field must adopt a multidimensional approach to validation, provenance, and governance. This includes scalable quality assessment, transparent licensing, ethical data stewardship, and privacy-preserving architectures.

## 3 Discussion

The "Knowledge Networks in Biomedical and Behavioral Research" meeting provided a forum for discussion about challenges, opportunities and future directions for a biomedical knowledge network. Our key observations and analyses of the Desiderata are summarized in **Figure 1**. While the Desiderata are presented in separate sections, we illustrate in a knowledge graph created by using terms from the Desiderata that their underlying principles intersect (**Figure 2**).

The Desiderata knowledge graph shows that the most highly connected and salient concepts relate to the unique challenges of representing the complex nature of the biomedical domain (e.g., heterogeneity, patient data), critical qualities of KGs (e.g., trustworthiness, transparency, reproducibility, interoperability), graph quality assessments (e.g., validation, provenance, privacy), and graph methods (e.g., inference, modeling, LLM, RAG). Indeed, one can argue that the virtues of a biomedical knowledge network lie in its trustworthiness. The concept of "trustworthy" can be described as epistemic components, relating to provenance and validation, and ethical components, such as privacy and security protections. As a final remark, we emphasize that as knowledge graphs become more central to biomedical AI, ensuring their integrity and social acceptability will be as important as ensuring their computational power.

## Acknowledgements

We wish to thank the NIH Knowledge Graph Working Group and the Office of Data Science Strategy (ODSS) for fostering this collaboration and the NIH organizing committee (Ishwar Chandramouliswaran, Mauricio Rangel Gomez, Chun-Ju Hsiao, Haluk Resat) for co-hosting this meeting. We thank the session panelists (Chris Bizon, Katy Börner, Licong Cui, Nick Evans, Jeffrey Grethe, Melissa Haendel, Bian Jiang, Chris Mungall, Jonathan Silverstein, Deanne Tayler, Chunhua Weng) for their panel presentations and discussion. We thank all meeting participants who contributed to the discussion.

## Author contributions

All authors equally contributed to writing this perspective.

## Conflict of interest

None declared.

## Data availability

No data is associated with this manuscript since it is a perspective.

## References

Brazma, A., P. Hingamp, J. Quackenbush, G. Sherlock, P. Spellman, C. Stoeckert, J. Aach, et al. 2001. "Minimum Information about a Microarray Experiment (MIAME)-toward Standards for Microarray Data." *Nature Genetics* 29 (4): 365–71.

Callahan, Tiffany J., Ignacio J. Tripodi, Adrianne L. Stefanski, Luca Cappelletti, Sanya B. Taneja, Jordan M. Wyrwa, Elena Casiraghi, et al. 2024. "An Open Source Knowledge Graph Ecosystem for the Life Sciences." *Scientific Data* 11 (1): 363.

Carbon, Seth, Robin Champieux, Julie A. McMurry, Lilly Winfree, Letisha R. Wyatt, and Melissa A. Haendel. 2019. "An Analysis and Metric of Reusable Data Licensing Practices for Biomedical Resources." *PloS One* 14 (3): e0213090.

CFDE. n.d. "Common Fund Data Ecosystem (CFDE)." Accessed June 19, 2025. https://commonfund.nih.gov/dataecosystem.

Collins, Gary S., Paula Dhiman, Constanza L. Andaur Navarro, Jie Ma, Lotty Hooft, Johannes B. Reitsma, Patricia Logullo, et al. 2021. "Protocol for Development of a Reporting Guideline (TRIPOD-AI) and Risk of Bias Tool (PROBAST-AI) for Diagnostic and Prognostic Prediction Model Studies Based on Artificial Intelligence." *BMJ Open* 11 (7): e048008.

"Data Management and Sharing Policy." n.d. Accessed June 19, 2025. https://sharing.nih.gov/data-management-and-sharing-policy.

DDKG. n.d. "Data Distillery KG." DDKG. Accessed June 19, 2025. https://dd-kg-ui.cfde.cloud/about.

Fecho, Karamarie, Gwênlyn Glusman, Sergio E. Baranzini, Chris Bizon, Matthew Brush, William Byrd, Lawrence Chung, et al. 2025. "Announcing the Biomedical Data Translator: Initial Public Release." *Clinical and Translational Science* 18 (7): e70284.

GDPR. 2018. "General Data Protection Regulation." GDPR. 2018. https://gdpr-info.eu/.

Haendel, Melissa A., James P. Balhoff, Frederic B. Bastian, David C. Blackburn, Judith A. Blake, Yvonne Bradford, Aurelie Comte, et al. 2014. "Unification of Multi-Species Vertebrate Anatomy Ontologies for Comparative Biology in Uberon." *Journal of Biomedical Semantics* 5 (1): 21.

ISO 27001. 2022. "ISO/IEC 27001:2022 Information Security, Cybersecurity and Privacy Protection — Information Security Management Systems — Requirements." ISO 27001. 2022. https://www.iso.org/standard/27001.

***Desiderata for a biomedical knowledge network***

KN meeting. 2025. "Meeting Agenda with Slide Presentations, 'Knowledge Networks in Biomedical and Behavioral Research' Meeting, Rockville, Maryland, May 28-29, 2025." KN Meeting. May 29, 2025. https://drive.google.com/file/d/1OZs5z0RWAi-dimu6kMimn8mx1p_KIBrT/view?usp=sharing.

Landi, Annalisa, Mark Thompson, Viviana Giannuzzi, Fedele Bonifazi, Ignasi Labastida, Luiz Olavo Bonino da Silva Santos, and Marco Roos. 2020. "The 'A' of FAIR – as Open as Possible, as Closed as Necessary." *Data Intelligence* 2 (1–2): 47–55.

Matentzoglu, Nicolas, Susan M. Bello, Ray Stefancsik, Sarah M. Alghamdi, Anna V. Anagnostopoulos, James P. Balhoff, Meghan A. Balk, et al. 2024. "The Unified Phenotype Ontology (UPheno): A Framework for Cross-Species Integrative Phenomics." *BioRxivorg*. https://doi.org/10.1101/2024.09.18.613276.

Patil, Ramesh S., Richard Fikes, Peter F. Patel-Schneider, Donald P. McKay, Timothy W. Finin, Thomas R. Gruber, and Robert Neches. 1992. "The DARPA Knowledge Sharing Effort: A Progress Report." In *Proceedings of the Third International Conference on Principles of Knowledge Representation and Reasoning*, 777–88. KR'92. San Francisco, CA, USA: Morgan Kaufmann Publishers Inc.

Pop, Mihai, Teresa K. Attwood, Judith A. Blake, Philip E. Bourne, Ana Conesa, Terry Gaasterland, Lawrence Hunter, et al. 2025. "Biological Databases in the Age of Generative Artificial Intelligence." *Bioinformatics Advances* 5 (1): vbaf044.

Proto-OKN. n.d. "Proto-OKN." Proto-OKN. Accessed June 19, 2025. https://www.proto-okn.net.

Putman, Tim E., Kevin Schaper, Nicolas Matentzoglu, Vincent P. Rubinetti, Faisal S. Alquaddoomi, Corey Cox, J. Harry Caufield, et al. 2024. "The Monarch Initiative in 2024: An Analytic Platform Integrating Phenotypes, Genes and Diseases across Species." *Nucleic Acids Research* 52 (D1): D938–49.

Resnik, David B. 1998. *The Ethics of Science: An Introduction*. Routledge. https://doi.org/10.4324/9780203979068.

Stear, Benjamin J., Taha Mohseni Ahooyi, J. Alan Simmons, Charles Kollar, Lance Hartman, Katherine Beigel, Aditya Lahiri, et al. 2024. "Petagraph: A Large-Scale Unifying Knowledge Graph Framework for Integrating Biomolecular and Biomedical Data." *Scientific Data* 11 (1): 1338.

UBKG. n.d. "Unified Biomedical Knowledge Graph (UBKG)." UBKG. Accessed June 19, 2025. https://ubkg.docs.xconsortia.org/.

Unni, Deepak R., Sierra A. T. Moxon, Michael Bada, Matthew Brush, Richard Bruskiewich, J. Harry Caufield, Paul A. Clemons, et al. 2022. "Biolink Model: A Universal Schema for Knowledge Graphs in Clinical, Biomedical, and Translational Science." *Clinical and Translational Science* 15 (8): 1848–55.

Whetzel, Patricia L., Natalya F. Noy, Nigam H. Shah, Paul R. Alexander, Csongor Nyulas, Tania Tudorache, and Mark A. Musen. 2011. "BioPortal: Enhanced Functionality via New Web Services from the National Center for Biomedical Ontology to Access and Use Ontologies in Software Applications." *Nucleic Acids Research* 39 (Web Server issue): W541-5.

# Figures

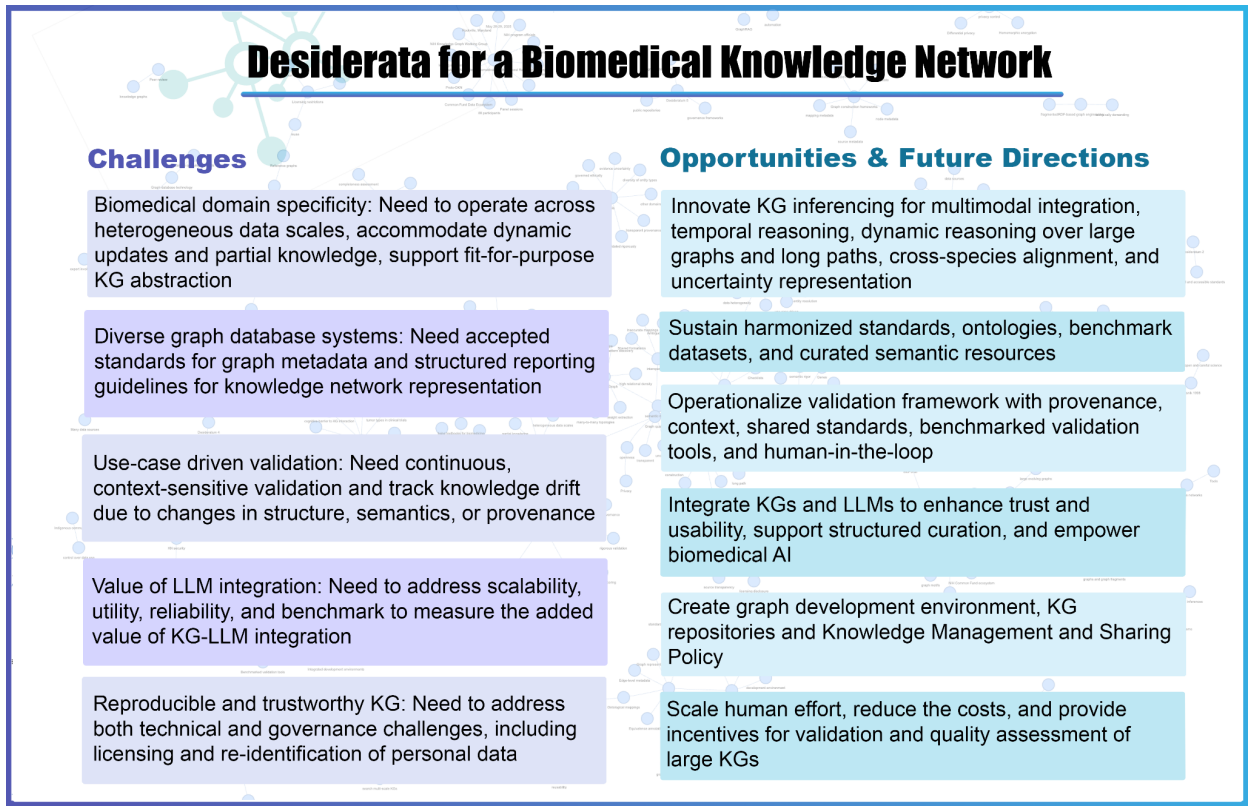

**Figure 1.** Desiderata for a biomedical knowledge network: challenges, opportunities and future directions.

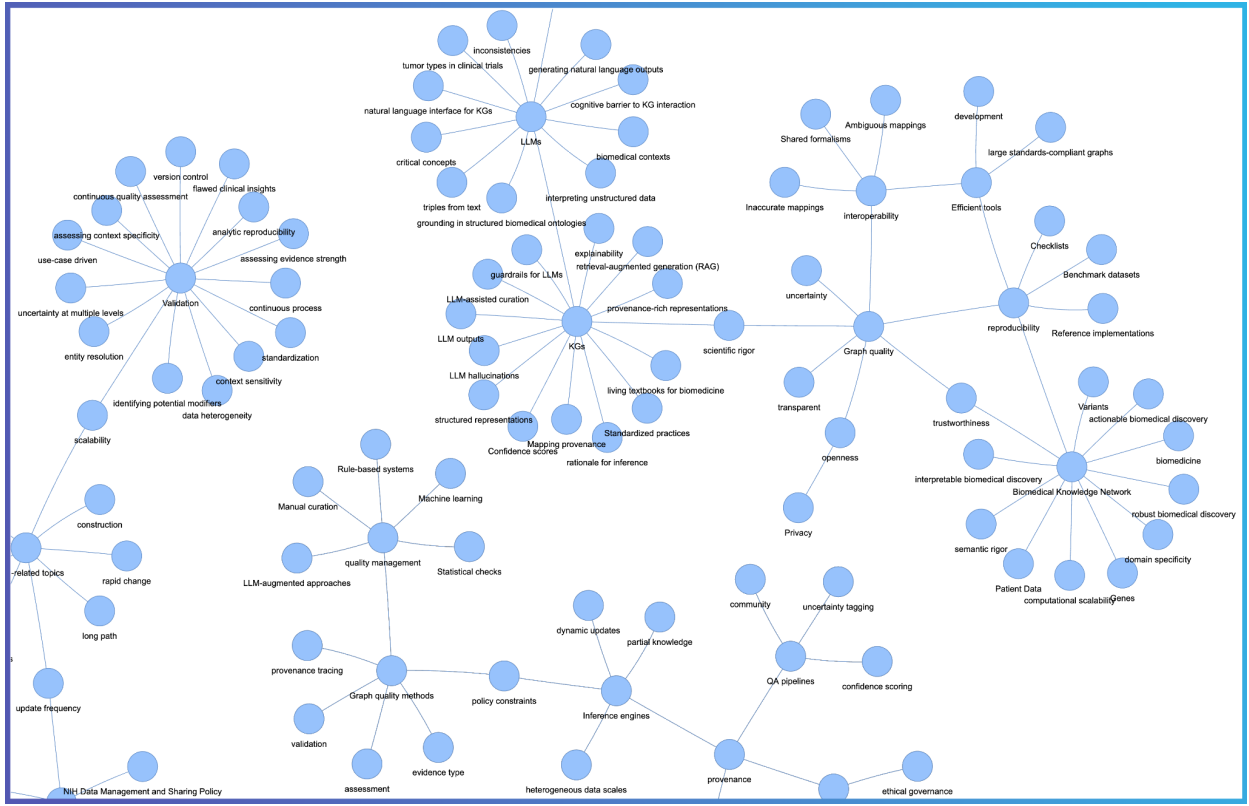

**Figure 2.** Convergent principles demonstrated as a knowledge graph from the Desiderata. The underlying graph is available at: https://biothings-data.s3.us-west-2.amazonaws.com/knmeeting/graph_desiderata.html.